\documentclass[11pt,preprint]{aastex}

\begin{document}

%
%
%

\title{Polarimetry with NICMOS}

%
%
%

\author{Dean C. Hines\altaffilmark{1}, Glenn Schneider\altaffilmark{2}}

\altaffiltext{1}{Space Science Institute, 4750 Walnut Street, Suite
205 Boulder, CO 80301} \altaffiltext{2}{Steward Observatory, The
University of Arizona, 933 N. Cherry Ave., Tucson, AZ 85721}

\begin{abstract}
NICMOS cameras 1 and 2 (with $\sim 43$ and $\sim 76$ mas/pixel,
respectively) each carry a set of three polarizing elements to provide
high sensitivity observations of linearly polarized light.  The
polarizers are bandpass limited and provide diffraction-limited
imaging in camera 1 at 0.8 - 1.3$\mu$m, and in camera 2 at
1.9-2.1$\mu$m.

The NICMOS design specified the intra-camera primary axis angles of
the polarizers to be differentially offset by $120\deg$, and with
identical polarizing efficiency and transmittance.  While this ideal
concept was not strictly achieved, accurate polarimetry in both
cameras, over their full (11\arcsec\ and $\sim$ 19.2\arcsec\ square)
fields of view was enabled through ground and on-orbit calibration of
the as-built and {\it HST}-integrated systems.

The Cycle 7 \& 7N calibration program enabled and demonstrated
excellent imaging polarimetric performance with uncertainties in
measured polarization fractions $\le 1\%$.  After the installation
of the NICMOS Cooling System (NCS), the polarimetric calibration was
re-established in Cycle 11, resulting in systemic performance
comparable to (or better than) Cycle 7 \& 7N.

The NCS era NICMOS performance inspired the development of an earlier
conceived, but non-implemented, observing mode combining high contrast
coronagraphic imaging and polarimetry in camera 2.  This mode was
functionally tested in the Cycle 7 GTO program, but without a detailed
characterization of the instrumental polarization induced by the
coronagraph, proper data calibration was not possible.  To remedy this
shortfall and to enable a new and powerful capability for NICMOS, we
successfully executed a program to calibrate and commission the
``Coronagraphic Polarimetry'' mode in NICMOS in Cycle 13, and the mode
was made available for GO use in Cycle 14.

We discuss the data reduction and calibration of direct and
coronagraphic NICMOS polarimetry.  Importantly, NICMOS coronagraphic
polarimetry provides unique access to polarized light near bright
targets over a range of spatial scales intermediate between direct
polarimetry and ground-based (coronagraphic) polarimetry using
adaptive optics.
\end{abstract}


\keywords{NICMOS, polarimetry, coronagraphy}


\section{Thermal Vacuum Tests}

The NICMOS polarimetry systems were characterized on the ground during
thermal vacuum tests using a light source that fully illuminated the
field of view with completely linearly polarized light (i.e., p(\%) =
100\%) and with position angles variable in 5$\deg$ increments.  The
polarizers in camera 1 and camera 2 are labeled by their nominal
position angle orientation: POLS* for camera 1 and POLL* for camera 2.
Camera 3 has three grisms that potentially act as polarizing elements,
and were characterized as well.  However, because the grisms reside in
the NIC3 filter wheel, they cannot be used with either the NIC1 or
NIC2 polarizers and are unsuitable for spectropolarimetry.

The primary results of these thermal vacuum tests include:

\begin{itemize}
\item Each polarizer in each camera has a unique polarizing
efficiency\footnote{Polarizer efficiency is defined as $\epsilon =
(S_{\rm par} - S_{\rm perp})/(S_{\rm par} + S_{\rm perp})$, where
$S_{\rm par}$ and $S_{\rm perp}$ are the respective measured signals
for a polarizer oriented parallel and perpendicular to the position
angle of a fully polarized beam.}, with POL120S having the lowest at
$\epsilon_{\rm POL120S}=48\%$.

\item The angular offsets between the polarizers within each filter
wheel differ from their nominal values of $120\deg$.

\item The instrumental polarization caused by reflections off the
mirrors and optical baffles in the optical train is small ($\le 1\%$).

\item The grisms indeed act as partial linear polarizers, with the
long wavelength grism (G206) producing the largest variation in
intensity ($\sim$5\%) for 100\% linearly polarized light.  If one
measures the equivalent width of an emission line against a polarized
continuum, the measurement will have different values depending on the
orientation of the spacecraft {\it w.r.t.} the position angle of the
polarized continuum.  This effect scales with the source polarization,
so should not be important for objects of low polarization (p(\%) $\le
10\%$).
\end{itemize}

\section{The Algorithm for Reducing NICMOS Polarimetry 
Observations}

The thermal vacuum results showed that the standard reduction
algorithm listed in basic optics text books would not work for NICMOS
data.  Instead, a more general approach was required (Hines, Schmidt
\& Lytle 1997; Hines 1998; Hines, Schmidt \& Schneider 2000; Hines
2002).

At any pixel in an image, the observed signal from a 
polarized source of total intensity $I$ and linear Stokes 
parameters $Q$ and $U$ measured through the $k^{th}$ 
polarizer oriented at position angle $\varphi_{k}$ is
\begin{equation}
S_k = A_kI + \epsilon_k(B_kQ + C_kU)~.
\end{equation}
Here,
\begin{equation}
A_k = \frac{t_k}{2}(1+l_k),~~ B_k = A_k\cos{2\varphi_k},~~ C_k 
= A_k\sin{2\varphi_k}~,
\end{equation}
$\epsilon_{k}$ is the polarizing efficiency, $t_k$ is the fraction of
light transmitted through the polarizer for a 100\% linearly polarized
input aligned with the polarizer axis, and $l_k$ is the ``leak'' --
the fraction of light transmitted through the polarizer (exclusive of
that involved in $t_k$) when the incident beam is polarized
perpendicular to the axis of the polarizer.  These quantities are
related under the above definitions, $\epsilon_k = (1-l_k)/(1+l_k)$.

This treatment can be shown to be equivalent to other approaches, once 
appropriate transformations are made (Mazzuca, Sparks, \& 
Axon 1998; see also Sparks \& Axon 1999).

The values of $t_{k}$ were determined initially by the filter
manufacturer from witness samples, and were not accurately remeasured
during thermal vacuum tests.  However, on-orbit observations of the
unpolarized and polarized standard stars enables refinement of these
numbers.  Adopted characteristics of the individual polarizers and
algorithm coefficients derived during and applicable to Cycle 7 \& 7N
are listed in Table~1 of Hines, Schmidt \& Schneider (2000), the 
Cycle 11 coefficients are listed in Hines (2002), and are
also available in the NICMOS instrument manual and the {\it HST} Data
Handbook.  The primary coefficients are also listed in Table 1 in 
this contribution.

After solving the system of equations (eq.  1) for the 
Stokes parameters at each pixel ($I, Q, U$), the percentage 
polarization ($p$) and position angle ($\theta$) are 
calculated in the standard way:
\begin{equation}
p = 100\% \times\frac{\sqrt{Q^{2}+U^{2}}}{I},~~~ \theta = 
\frac{1}{2}{\rm tan}^{-1}\left( \frac{U}{Q} \right)~.
\end{equation}
Note that a 360$\deg$ arctangent function is assumed.

This algorithm has been tested by the NICMOS Instrument Definition
Team (IDT) and by the Space Telescope Science Institute (STScI) on
several data sets.  An implementation has been developed by the IDT,
and integrated into a software package written in IDL. The package is
available through the STScI
website\footnote{http://www.stsci.edu/instruments/nicmos/ISREPORTS/isr\_99\_004.pdf}
and is described by Mazzuca \& Hines (1999).  An update with improved
graphical user interface will be available for Cycle 15.

\section{The Cycles 7, 7N \& 11 Polarimetry Characterization Programs}

Observations of a polarized star (CHA-DC-F7: Whittet et al.  1992) and
an unpolarized (null) standard (BD + 32û 3739: Schmidt, Elston \&
Lupie 1992) were obtained with NIC1 and NIC2 in Cycle 7 \& 7N (CAL
7692, 7958: Axon PI) and in Cycle 11 (CAL 9644: Hines PI).  Data were
obtained at two epochs such that the differential telescope roll
between observations was $\approx 135\deg$.  This removes the
degeneracy in position angle caused by the pseudo-vector nature of
polarization.  The second epoch observations in Cycle 7N and all of
the observations in Cycle 11 used a four position, Òspiral-ditherÓ
pattern with 20.5 pixel (NIC1) and 30.5 pixel (NIC2) offsets to
improve sampling and alleviate the effects of bad pixels, some
persistence, and other image artifacts.  This is also the recommended
observing strategy for all NICMOS, direct imaging polarimetry
programs.  While no dither pattern was used during the first epoch of
Cycle 7, the data do not appear to suffer significantly from
persistence.  Observations of the highly polarized, proto-planetary
nebula CRL 2688 (Egg Nebula) were also obtained to test the
calibration over a large fraction of the field of view.

The observations were processed through the CALNICA pipeline at STScI
using the currently available reference files.  Aperture photometry
was used to measure the total flux density from the stars in
instrument units (DN/s) for each polarizer.  The Stokes parameters
were then constructed using equation (1).  Since the thermal vacuum
tests showed that the imaging optics themselves have little effect on
the observed polarization, any measured polarization in the null
standard was attributed the $t_{k}$ term in the algorithm.  The
algorithm was applied to the polarized standard stars to check both
the percentage polarization and the position angles.

The higher, yet more stable, operating temperature provided by the NCS
and the three year dormancy of NICMOS may contribute to changes in the
metrology of the optical system.  Therefore, a program to
re-characterize the polarimetry optics in Cycle 11 has been developed
(NIC/CAL 9644: Hines).

The basic design of the Cycle 11 program follows the strategy
undertaken in Cycle 7 \& 7N, relying on observations of polarized and
unpolarized standard stars as well as the proto-planetary nebula CRL
2688 (Egg Nebula).  The Egg Nebula was only observed at a single epoch
as a direct comparison with observations from Cycle 7 \& 7N (ERO 7115:
Hines; Sahai et al.  1998; Hines, Schmidt \& Schneider 2000), and to
evaluate any gross discrepancies across the field of view.
Observations of the unpolarized standard star (p(\%)$_{\rm intrinsic}
= 0\%$, by definition) processed with the Cycle 7 \& 7N algorithm
coefficients yield $p_{\rm NIC1} = 0.7\% \pm 0.2\%, \theta_{\rm NIC1}
= 74\deg$ and $p _{\rm NIC2} = 0.7\% \pm 0.3\%, \theta_{\rm NIC2} =
73\deg$.  This suggests that the system changed.  The observations of
the polarized standard star is also larger ($\Delta p \approx 2\%$) in
NIC1 compared with the measurements of Cycle 7 \& 7N, which themselves
were in excellent agreement with ground-based measurements (Hines,
Schmidt \& Schneider 2000).

Observations of the Egg Nebula were also analyzed with the Cycle 7 \&
7N coefficients.  As for the polarized standard star, the results for
the Egg Nebula suggest that the polarimetry system has changed
slightly\footnote{The polarization structure of the Egg is not
expected to change over the 5 year period between observations even
though the object is known to show photometric variations.}, again by
about 2\% in p\%.

We rederived the coefficients assuming that any measured polarization
in the unpolarized (Q=U=0) standard was attributed to the t$_{k}$
term in the algorithm (i.e. the variable term in the A$_{k}$ 
coefficient).  The new coefficients are listed in Table 1.

We note that we are embarking on a new study of all of the NICMOS
direct imaging polarimetry calibration data in an attempt to improve
the calibration to better than 0.3\% (Batcheldor et al.  2006).  This
study was prompted by reports of a $\sim 1.5\%$ systematic residual
polarization signal in some objects (e.g., Ueta et al.  2005).
Indeed, Batcheldor et al.  find a residual excess polarization for the
ensemble of (un)polarized standard stars of p(\%) $\approx 1.2\%$,
which behaves as a constant ``instrumental'' polarization that can be
subtracted (in Stokes parameters) from the observations.  While
insignificant for highly polarized objects, this residual will affect
observations of low polarization objects such as active galactic
nuclei.

Due to the non-ideal polarimetry optical system, estimation of the
uncertainties in the percentage polarization require a slighlty more
sophisticated analysis, especially for NIC 1.  Monte Carlo simulations
of the uncertainties based on the NICMOS performance are given in
Hines, Schmidt \& Schneider (2000), and an analytical approach is
given in Sparks \& Axon (1999).  In particular, the usual assumption
that $\sigma_{q} = \sigma_{u}$ is not valid for NIC 1.

\section{Coronagraphic Polarimetry with NICMOS}
Coronagraphic polarimetric observations of the same unpolarized and
polarized standard stars that were used in Cycle 7, 7N (CAL 7692,
7958: Axon PI) and Cycle 11 (CAL 9644: Hines PI), and observations of
the face-on circumstellar disk TW Hya were each obtained at two epochs
sufficiently spaced in time to permit large differential rolls of the
spacecraft (i.e., field orientations {\it w.r.t.} the HST optics and
NICMOS polarizers).  At each epoch, coronagraphic polarimetric imaging
was carried out at two field orientations differing by $29.9\deg$.
Following standard NICMOS Mode-2 target acquisitions, the acquired
targets were observed through each of the three ``Long'' wavelength
polarizers in Camera 2.  The intra-visit repeats for the standard
stars were designed to check both for repeatability (image stability)
and possible image persistence (none of consequence were found).  TW
Hya was exposed in only two repeats for each polarizer.

To facilitate commonality in data processing, all coronagraphic imaging
was done with STEP16 multiaccum sampling, though the number of samples
varied between 10 to 12 to best fill the orbits for each of the
targets.  The data were instrumentally calibrated in an APL-based
analog to the STSDAS CALNICA task using on-orbit derived calibration
reference files suitable for these observations.  Following the
creation of count-rate images individual bad pixels were replaced by
2D weighted Gaussian interpolation (r=5 weighing radius) of good
neighbors, and ``horizontal striping'' associated with heavily exposed
targets was characterized and removed by median-collapse subtraction.

\subsection{PSF-subtracted Images}

The image data for this program were obtained to investigate and
demonstrate the further suppression of instrumentally scattered light
via the camera 2 coronagraph (compared to non-coronagrahic
polarimetry) and thus effect a reduction in the detection floor for a
polarized signal in the diffuse wings of the (supressed) scattering
component of the occulted PSF core.  For this purpose, the data
obtained were not optimized to enable coronagraphic PSF subtraction
(see, e.g., Fraquelli et al.  2004).  Imaging coronagraphy under the
pass band Camera 2 1.9-1.2$\mu$m polarizers will be less effective
than under the 1.4-1.8$\mu$m F160W (H band) filter.  Imaging
coronagraphy of our calibration science target, the very-near face-on
TW Hya circumstellar disk under the F160W filter was discussed in
Weinberger et al.  (2002).  The smaller inner working angle of the
coronagraph with polarizers (caused by the longer wavelength
pass-band), while fully enclosing the PSF core under the polarizers
permits light from the first Airy ring to escape the coronagraphic
image plane mask.

Both the unpolarized standard BD +32 -3739 and the polarized standard
CHA-DC-F7 were selected to establish ``truth'' {\it w.r.t.} the
instrumental polarization (and ultimately sensitivity) with
polarimetric coronagraphy.  They were not chosen to serve as
coronagraphic reference PSF's for subtraction from TW Hya, and both
calibration stars are significantly fainter than TW Hya.
Additionally, the field of the unpolarized standard has a significant
number of fainter stars near the target, unknown prior to the
coronagraphic imaging resulting from this program.  This is not a
problem for the coronagraphic polarimetric calibration/validation, but
as such that additionally makes it a poor choice as a PSF subtraction
template star.

Nonetheless, DIRECT imaging of circumstellar dust from the Camera 2
polarizer set derived total intensity images is demonstrated (but not
comprehensively in this summary) using this ``template'' PSF. The
depth of exposure of BD+32 is insufficient to reveal the outer part of
the TW Hya disk.  The flux density rescaling of BD+32 is suggested as
3.48 from the 2MASS H and K magnitudes and the passband of the Camera
2 polarizers.  The disk persists varying the DB+32 intensity over the
full range of TW Hya:BD+32 intensities suggested in H and K, though
very significant under and over subtraction residuals appear at the
extreme of that range in the diffraction spikes, suggesting the a
prior ratio was very close to ``correct''.

\subsection{Polarimetry Ð Comparison Between the Unpolarized Star
and TW Hya}

After reduction of the images in the three different polarizers, the
images were processed through the NICMOS polarimetry reduction
algorithm (Eq.  1'' Hines, Schmidt \& Schneider 2000) to produce
images of the polarization position angle ($\theta$), the percentage
polarization, and the polarized intensity ($= p\times I$).  To
consider the detection of polarized flux to be ``well determined'' we
demand the polarized intensity in any pixel be at least 5-sigma above
the median of the background.  For this data set we choose a
background region far from the target, and away from the diffraction
spikes, where no contribution to the background from the target flux is
seen.  With this, the centosymmetric polarization field for the
near-face on TW Hya disk is obvious, and highly repeatable, in all
four field orientations of the TW Hya observations.

While some residual instrumental polarization seems apparent, the lack
of obvious structure in the position angle image of the unpolarized
standard star is dramatically different than that of the TW Hya
position angle image; the TW Hya image shows the classic ``butterfly''
pattern for centrosymmetric polarization cased by scattering off a
face-on disk.  Furthermore, the median polarization is about 2\% per
resolution element for the ``unpolarized'' star, compared with $\sim
10-15\%$ per resolution element for TW Hya.  The results for TW Hya
are shown in Figure 1.  The measured polarization in the disk is
consistent with measurements made from the ground with Adaptive
Optical systems (D. Potter, private comm., and Hales et al. 2005).

\subsection{Continued Analysis}

Our analysis so far has shown that the mode is viable for future
Cycles.  We have sufficient data in hand to evaluate:

\begin{itemize}
\item The ability to remove the small instrumental polarization
signature in Stokes parameter space.

\item  The spatial correlation of this instrumental polarization as a
function of radius and azmuth with respect to the coronagraphic hole.

\item The polarization sensitivity as a function of radius from the
coronagraphic hole.  This analysis is limited by the exposure
depth of the unpolarized standard.

\end{itemize}

\subsection{Recommendations}
Based upon our preliminary analysis, we recommend that NICMOS
coronagraphic polarimetry be released as an available mode for future
Cycles.  We note that the same rules that apply to direct
coronagraphic imaging apply to this mode as well.  In particular, the
fine guidance necessary to keep the target well centered in the hole
is necessary.  In addition, we recommend that observations be obtained
at two-roll angles to interpolate over bad pixels and develop an in
situ flat field correction near the coronagraphic hole.  This may be 
difficult to achieve in with two-gyro mode.

We cannot yet tell if the current calibration data from Cycle 12 will
be sufficient for GOs.  At present, we would urge that GOs observe
their target plus an unpolarized standard star, each at sufficient
depth to obtain similar S/N in each object in each polarizer Ð thus a
minimum of two orbits for a single target program.  However, a single,
well exposed unpolarized standard star should be sufficient for a
multi-target science program.

\section{Two Gyro Mode}
Based upon the measured jitter in two-gyro mode, there
should be no degradation of the direct polarimetry mode.  Tests of
coronographic imaging also suggest minimal degradation (Schneider et
al.  2005, ISR2005-001).  The primary impact will be the inability to
obtain images at two spacecraft roll angles within a single orbit.

\section{Summary}
A wide variety of astronomical objects have been examined with NICMOS
direct imaging polarimetry including Active Galaxies (Capetti et al.
2000; Tadhunter et al.  2000; Simpson et al.  2002), evolved stars and
proto-planetary nebulae (Sahai et al.  1998; Weintraub et al.  2000;
Schmidt, Hines \& Swift 2002; King et al.  2002; Su et al.  2003; Ueta
et al.  2005), and young stellar objects (Silber et al.  2000; Meakin,
Hines \& Thompson 2005).  NICMOS continues to provide high fidelity,
high spatial resolution imaging polarimetry.  Combined with the
polarimetry mode of the Advanced Camera for Surveys (ACS), {\it HST}
provides high resolution imaging polarimetry from $\sim 0.2 -
2.1\mu$m.  With the newly commissioned coronagraphic polarimetry mode,
NICMOS provides further unique access to polarized light near bright
targets over a range of spatial scales intermediate between direct
polarimetry and ground-based (coronagraphic) polarimetry using
adaptive optics.

An electronic version of the presentation given by D.C. Hines at the 
workshop can be found at the following URL:

http://www.stsci.edu/ts/webcasting/ppt/CalWorkshop2005/DeanHines102805.ppt

\begin{figure}
\epsscale{0.80}
\plotone{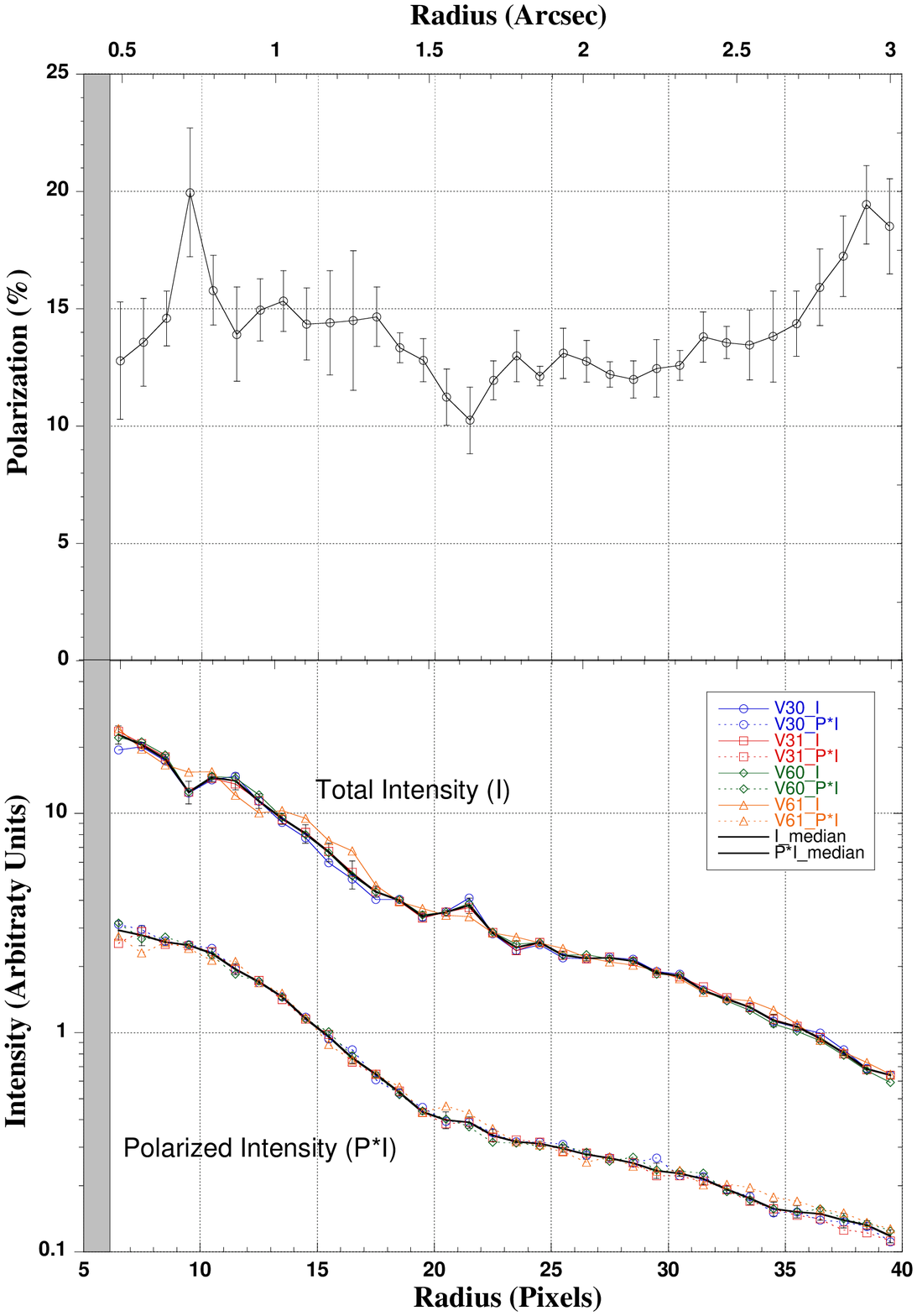}
\caption{Coronagraphic polarimetry of TW Hya.  The top panel shows the
percentage polarization.  The bottom panel shows the azmuthally
averaged total intensity and the polarized intensity.}
\label{example-fig-1}
\end{figure}

\begin{deluxetable}{lccccccccl}
\tablewidth{0pt}
\tablecaption{Characteristics of the NICMOS Polarizers \& Coefficients to Eq. (1)} 
\tablehead{\colhead{Filter} & 
\colhead{$\varphi_{k}$} & \colhead{$\epsilon_{k}$} 
& \colhead{t$_{k}$(Pre-NCS)\tablenotemark{a}} &
\colhead{t$_{k}$(NCS)\tablenotemark{a}} & \colhead{Comments}}
\startdata
POL0S  &  1.42  & 0.9717 & 0.7760 & 0.7760  & Ghosts \\
POL120S& 116.30 & 0.4771 & 0.5946 & 0.5934  & Weak ghosts \\
POL240S& 258.72 & 0.7682 & 0.7169 & 0.7173  & Ghosts \\
       &        &        &        &         &        \\ 
POL0L  &   8.84 & 0.7313 & 0.8981 & 0.8779  & \ldots \\
POL120L& 131.42 & 0.6288 & 0.8551 & 0.8379  & \ldots \\
POL240L& 248.18 & 0.8738 & 0.9667 & 0.9667  & \ldots \\
\enddata
\tablenotetext{a}{Derived from on-orbit observations of the 
unpolarized standard star BD$+32\deg$3739 (Schmidt, Elston \& 
Lupie 1992).}
\end{deluxetable}

\end{document}